\documentclass[a4paper,11pt]{article}
\usepackage{jheppub} 
\usepackage{amssymb}
\usepackage{epsfig}
\usepackage{graphicx}
\usepackage[T1]{fontenc}

\def \lm  {\lambda}

\def \be {\begin{equation}}
\def \ee {\end{equation}}
\def \barr {\begin{array}{lc}}
\def \earr {\end{array}}

\def \lf  {\left (}
\def \rt  {\right )}
\def \pa  {\partial}
\def \f  {\frac}

\def \lm {\lambda}

\title{\boldmath Rindler-AdS/CFT}

\author[a]{Maulik Parikh}
\author[b]{and Prasant Samantray}

\affiliation[a]{Beyond: Center for Fundamental Concepts in Science\\
Arizona State University, Tempe, Arizona 85287, USA}
\affiliation[b]{Department of Physics\\ 
Arizona State University, Tempe, Arizona 85287, USA}

\emailAdd{maulik.parikh@asu.edu}
\emailAdd{prshkumar@gmail.com}

\abstract{In anti-de Sitter space a highly accelerating observer perceives a Rindler horizon. The two Rindler wedges in $AdS_{d+1}$ are holographically dual to an entangled conformal field theory that lives on two boundaries with geometry $\mathbb{R} \times H_{d-1}$. For AdS$_3$, the holographic duality is especially tractable, allowing quantum-gravitational aspects of Rindler horizons to be probed. We recover the thermodynamics of Rindler-AdS space directly from the boundary conformal field theory. We derive the temperature from the two-point function and obtain the Rindler entropy density precisely, including numerical factors, using the Cardy formula. We also probe the causal structure of the spacetime, and find from the behavior of the one-point function that the CFT ``knows'' when a source has fallen across the Rindler horizon. This is so even though, from the bulk point of view, there are no local signifiers of the presence of the horizon. Finally, we discuss an alternate foliation of Rindler-AdS which is dual to a CFT living in de Sitter space.}

\begin{document}

\maketitle
\flushbottom
\setcounter{page}{2}
\setcounter{footnote}{0}


\bigskip
\bigskip

\section{Introduction}

Rindler space, the portion of Minkowski space with which an observer undergoing constant acceleration can interact, is perhaps the simplest spacetime with a horizon. As the near-horizon limit of all nonextremal black holes and an example of a spacetime with an observer-dependent horizon, Rindler space has been much studied. Nevertheless, most of the literature on the subject has treated Rindler space using the techniques of quantum field theory in curved spacetime, whereas it is now recognized that many of the most interesting problems of horizon physics are not accessible with those techniques. Instead one would like to be able to study Rindler space in a theory of quantum gravity. This has not been done for the simple reason that a tractable theory of quantum gravity in asymptotically flat space does not presently exist.

Fortunately, a tractable theory of quantum gravity in anti-de Sitter space does exist: it is defined by the AdS/CFT correspondence. This motivates us to consider accelerating observers not in Minkowski space but in AdS space. Observers in anti-de Sitter space with suitably high proper acceleration (compared with the AdS length scale) have acceleration horizons; Rindler-AdS space is thus the portion of anti-de Sitter space that such observers can interact with. The purpose of this paper is to set up a holographic duality between Rindler-AdS space and a boundary conformal field theory, and to then use that correspondence to investigate quantum-gravitational aspects of Rindler-AdS space. It is worth emphasizing that Rindler-AdS space is a particularly advantageous spacetime for studying the quantum gravity of horizons. Unlike eternal black holes in AdS, Rindler-AdS has no singularities where bulk physics breaks down. And unlike flat Rindler space, the existence of a dual conformal field theory is assured; indeed, in the case of AdS$_5$ it is known to be ${\cal N} = 4$ super Yang-Mills theory. Thus in principle one has all the tools necessary to study event horizons in a theory of quantum gravity.

While Rindler-AdS space in general dimensions has been described and studied previously, the real power of the AdS/CFT correspondence can be brought to bear when the bulk spacetime dimension is three. For that special case, the boundary theory becomes a two-dimensional CFT living in Minkowski space, with all the ensuing advantages. In particular, the two-point function can be calculated explicitly and the Rindler entropy density can be derived from the Cardy formula. The result matches the Bekenstein-Hawking entropy density of the Rindler horizon precisely, including numerical factors. Even more interestingly, one can probe the causal structure of the spacetime. Remarkably, we find that the boundary theory ``knows'' when a source has fallen past the Rindler horizon even though, from a bulk point of view, there are no local invariants that mark the presence of the event horizon.

This paper is organized as follows. In Section 2, we present the classical geometry of Rindler-AdS space. In Section 3, we quickly review Rindler-AdS thermodynamics. Section 4 describes the boundary theory and contains our main results. The results of the paper are as follows. We calculate the bulk-boundary propagator and the two-point correlation function of operators in the boundary theory. Specializing to AdS$_3$, we show that the Cardy formula precisely reproduces the Bekenstein-Hawking entropy density, including the numerical coefficient, both for nonrotating and rotating Rindler-AdS space. We then discuss the relation between Rindler-AdS space and AdS black holes. Next, we turn to perhaps our most interesting derivation. We consider a source that falls freely into the Rindler horizon. By calculating the one-point function of a boundary operator, we show that a ``boundary theorist'' can tell whether the source has fallen across the horizon. This is the main result of the paper. In Section 5, we consider an alternate foliation of Rindler-AdS in which the boundary conformal field theory lives in de Sitter space. We briefly discuss some subtleties of this variant of Rindler-AdS/CFT. We summarize and conclude in Section 6 with some remarks about directions and puzzles suggested by Rindler-AdS/CFT.

\section{The Geometry of Rindler-AdS Space}

We would like to cover anti-de Sitter space in the Rindler coordinates natural to an accelerating observer. AdS$_{d+1}$ can conveniently be described using embedding coordinates of $d+2$-dimensional Minkowski space with two time-like directions:
\be
-\left ( X^0 \right )^2 + \left ( X^1 \right )^2+ ... + \left ( X^d \right )^2 - \left ( X^{d+1} \right )^2 = - L^2 \; . \label{embed}
\ee
Here the AdS curvature scale is $L$ and the $O(2,d)$ isometry group is manifest. In the embedding space, a Rindler observer is one whose Hamiltonian is a boost generator. It was shown in an elegant paper \cite{deserlevin} that both acceleration and ``true'' horizons in an Einstein space (such as say Schwarzschild, de Sitter, or anti-de Sitter) can be regarded as Rindler horizons in a higher-dimensional flat embedding space. The Hawking or Unruh temperature detected by observers in the lower-dimensional space can be obtained directly from accelerating trajectories in the embedding space \cite{Jorge}.\footnote{This is because the response of Unruh detectors depends on the Wightman function which in turn depends only on geometric invariants (constructed out of bi-vectors) that can just as well be computed in the embedding space.} In particular, Rindler observers in AdS are also Rindler observers in the embedding Minkowski space \cite{deserlevinads}. 

\begin{figure}[hbtp]
 \centering
  \epsfysize=6 cm
  \epsfbox{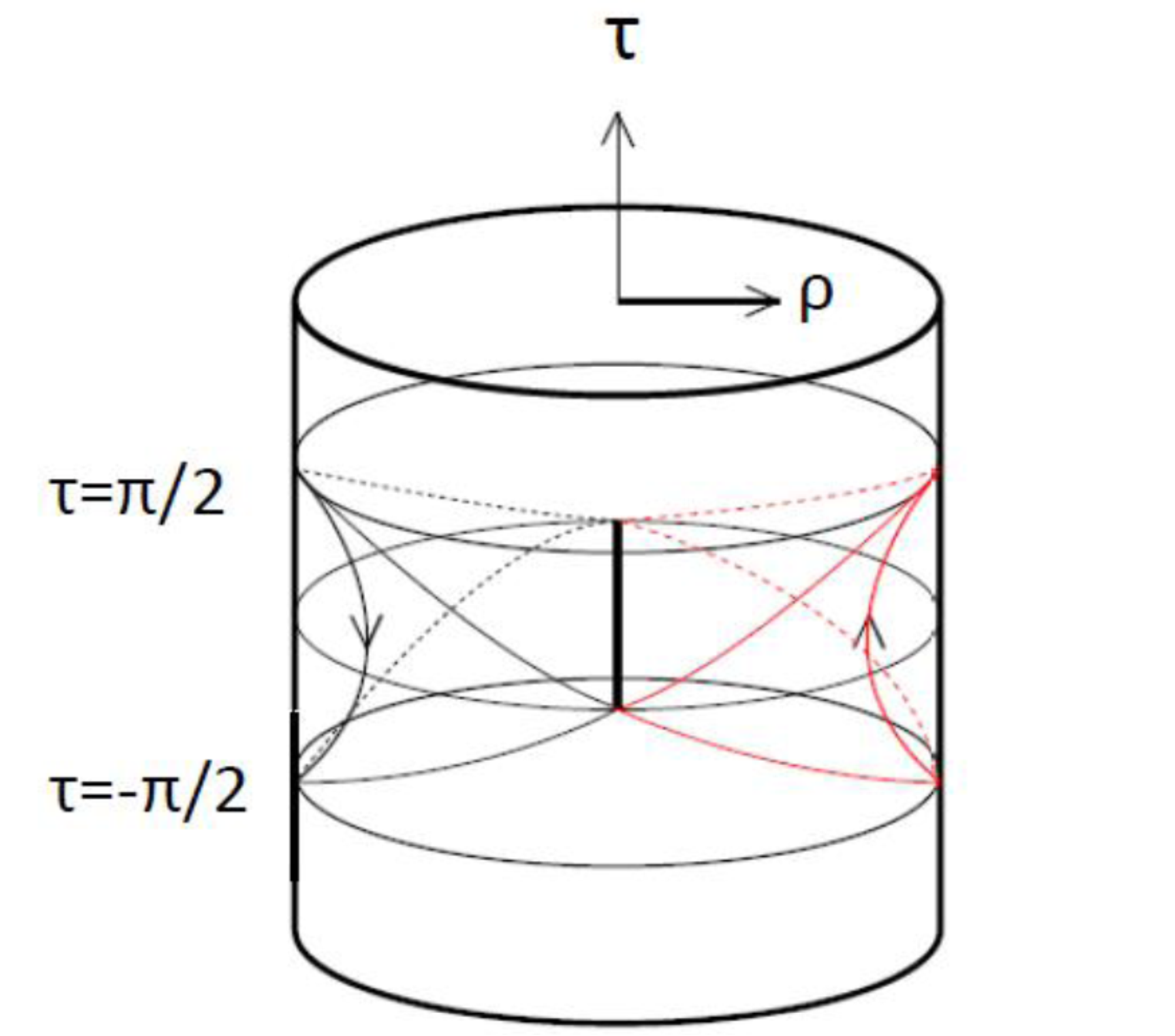}
 \centering
 \caption{Geometry of Rindler-AdS$_{d+1}$ space. A surface of constant $\xi$ is a $\mathbb{R} \times H_{d-1}$ hypersurface. $\tau$ and $\rho$ are the time and radius in global coordinates; except at $\rho = 0$ each point in the interior corresponds to a $S^{d-2}$. The Rindler-AdS region extends only up to $\tau = \pm \pi/2$ at the boundary of AdS. The arrow on the right points in the direction of $\partial_t$, whose orbits are a Rindler observer's worldline; the arrow is reversed for the antipodal observer.  One copy of the CFT lives on the boundary within the region shown in red.}
 \label{fig:geometry}
 \end{figure}

Consider then a Rindler observer in $d+2$-dimensional Minkowski space (with two time directions) uniformly accelerating in the $X^1$ direction:
\be
X^0 = {\xi} \sinh (t/ L) \qquad X^1 = {\xi} \cosh (t/ L) \; . \label{RindlerObs}
\ee
Here, instead of choosing an arbitrary acceleration parameter $g$ (as one does in flat Rindler space), we have used the existence of the AdS scale $L$ to rescale the time coordinate such that $g$ is replaced by $1/L$; since $g$ is unphysical, there is no loss of generality. Choosing the rest of the coordinates (see appendix) such that the embedding equation (\ref{embed}) is satisfied, the Rindler-AdS metric becomes:
\be
ds^2 =  - (\xi / L)^2 dt^2 + {d\xi^2 \over 1 + (\xi/L)^2} +  (1 + (\xi/L)^2) \left [ d\chi^2 + L^2 \sinh^2 (\chi/L)  d \Omega_{d-2}^2 \right ] \; . \label{RindlerAdSmetric}
\ee
This line element describes AdS in Rindler coordinates. These coordinates cover the part of the hypersurface (\ref{embed}) with $\left (X^1\right)^2 - \left (X^0\right )^2 > 0$ and $X^1, X^{d+1} > 0$. The above metric has been discussed in \cite{Emparan, Myers, Lowe, Lowe2, Vanzo, Raamsdonk} in various other contexts. Note that the constant-$\xi$ hypersurfaces are of the form $\mathbb{R} \times H_{d-1}$. These are the hypersurfaces on which the boundary CFT will be defined. The coordinate time $t$ parameterizes the worldline of an accelerating observer in AdS. Indeed, as the AdS curvature scale diverges, so that $\f{\xi}{L} \rightarrow 0$ and $L^2 \sinh^2 (\chi/L) \rightarrow \chi^2$, we recover
\be 
ds^2 =  - (\xi / L)^2 dt^2 + {d\xi^2} +  d\chi^2 + \chi^2  d \Omega_{d-2}^2  \; ,
\ee
which is just the line element of standard (i.e. flat) $d+1$-dimensional Rindler space. To understand the global properties of Rindler-AdS space, it is useful to consider AdS$_{d+1}$ in global coordinates (see appendix) for which the line element is
\be
ds^2 = - (1+ (\rho/L)^2) d\tau^2 + {d\rho^2 \over  1+ (\rho/L)^2} + \rho^2 d \Omega_{d-1}^2 \; . \label{global}
\ee
The global coordinates can then be expressed in terms of the Rindler-AdS coordinates as
\begin{eqnarray}
\rho^2 & = & \xi^2 \left [\cosh^2 (\chi /L) +  \sinh^2 (t/ L) \right ] + L^2 \sinh^2 (\chi /L) \nonumber \\
\tan \psi & = & {\sqrt{\xi^2 + L^2} ~ \sinh (\chi /L) \over \xi \cosh (t/ L)} \nonumber \\
\cos^2 (\tau/L) & = & {({\xi^2 + L^2}) \cosh^2 (\chi /L) \over \xi^2 \left [\cosh^2 (\chi /L) +  \sinh^2 (t/ L) \right ] + L^2 \cosh^2 (\chi /L)} \; .
\end{eqnarray}
Here $\psi$ is the polar angle on the $S^{d-1}$, which we have explicitly separated from the angles on the $S^{d-2}$. The angles $\theta_i, \phi$ on the $S^{d-2}$ are the same in both coordinate systems. In particular, the last equation indicates that the time-slice $t=0$ corresponds to $\tau = 0$. At other times, the constant-time slices of $t$ are tilted with respect to the constant-time slices of $\tau$. Furthermore, with the other coordinates held fixed, $\tau \rightarrow \pm {\pi \over 2}$ as $t \rightarrow \pm \infty$. Our Rindler coordinates therefore cover a finite interval of global time. This is illustrated in Figure \ref{fig:geometry}.

Since many of our calculations will be done in three dimensions, let us briefly consider that special case. The metric for Rindler-AdS$_3 $ is
\be
ds^2 = -{\xi^2\over L^2} dt^2 + {d\xi^2 \over 1 + {\xi^2 \over L^2}} + \left (1 + {\xi^2 \over L^2} \right ) d\chi^2 \; . \label{3dimRAdS} 
\ee
Its asymptotic behavior near the AdS boundary is given by
\be
ds^2 \rightarrow  {L^2 d\xi^2 \over \xi^2} + \frac{\xi^2}{L^2} \left (-dt^2 + d\chi^2 \right ) \; .
\ee
We see that, unlike in higher dimensions, the metric on a constant-$\xi$ hypersurface is conformal to Minkowski space. Moreover, as $\xi \to \infty$, the transformation $ \xi \rightarrow \gamma\xi$ and $ (\chi,t)\rightarrow \gamma^{-1} (\chi,t)$ is the usual scale-radius duality, and is manifestly an isometry of the asymptotic metric.

Another feature unique to three-dimensional anti-de Sitter space is the existence of a kind of rotating Rindler space \cite{RJAdS}:
\be
ds^2 = -\left( (\xi/L)^2 (1 - \beta^2) - \beta^2 \right) dt_r^2  - 2\beta dt_r \, d\chi_r + {d\xi^2 \over 1 + (\xi/L)^2} + \left(1 + (\xi/L)^2(1 - \beta^2) \right) d\chi_r^2  \; . \label{Rotating}
\ee
Here $-\infty < \chi_r < \infty$ and $\beta$ is a rotation parameter with $-1 \leq \beta \leq 1$. Both rotating and nonrotating Rindler-AdS space are of course a piece of anti-de Sitter space just as flat Rindler space is a piece of Minkowski space. In fact, even globally the portion of the spacetime covered by the coordinates above is identical to that covered by nonrotating Rindler 
coordinates. The diffeomorphism
\be
t \to t_r - \beta \chi_r	\qquad \chi \to \chi_r - \beta t_r \label{NR-to-R}
\ee
maps one spacetime to the other. In that sense, rotating Rindler space is classically the same spacetime as nonrotating Rindler space. However, the Hamiltonians for nonrotating and rotating Rindler space are not related by AdS isometries (they are in different conjugacy classes of the AdS isometry group) and the corresponding vacuum states (``$\beta$-vacua'') of scalar field theory are particle-inequivalent  \cite{RJAdS}. That is, the $\beta$-vacuum annihilated by the Hamiltonian that generates a rotating Rindler time appears to the nonrotating Rindler observer as an excited state populated with particles. Interestingly, rotating Rindler-AdS space possesses not only an observer-dependent event horizon but even an observer-dependent ergosphere at $\xi/L = \beta/\sqrt{1-\beta^2}$ \cite{RJAdS}.

\section{Thermodynamics of Rindler-AdS}

Contrary to the situation in flat space, the temperature seen by an observer moving with constant acceleration in curved spacetime is not always proportional to the proper acceleration. Rather, the general formula relating proper acceleration $a$ and local temperature in $(A)dS_{d+1}$ from \cite{deserlevinads} is
\be
T_{\rm local} = {1 \over 2 \pi} \sqrt{{2 \Lambda \over d(d-1)} + a^2} = {1 \over 2 \pi} a_{\rm embed} \; ,\label{tempgeneral}
\ee
where $a_{\rm embed}$ is the proper acceleration of the Rindler observer in the flat embedding space. This agrees for example with the fact that even a geodesic observer ($a = 0$) in de Sitter space sees a temperature. In AdS, there is a critical acceleration ($a_c = 1/L$) before the observer detects thermality. Observers at the critical acceleration see zero-temperature extremal horizons. Observers with lower acceleration do not have horizons. For example, an observer at a constant nonzero global radial coordinate $\rho$, moving in the direction of $\partial_\tau$, has a constant nonzero acceleration but nevertheless does not measure a temperature. Such an observer moves vertically up the Penrose diagram and has no horizons. From the embedding point of view, sub-critical acceleration trajectories correspond to spacelike trajectories in the higher-dimensional space and therefore do not give an Unruh temperature.

Consider then a Rindler-AdS observer at constant $\xi$. The proper acceleration of such an observer is
\be
a^2 = {1\over \xi^2} + {1 \over L^2} \; . \label{accel}
\ee
Inserting (\ref{accel}) into (\ref{tempgeneral}) we get
\be
T_{\rm local} = {1 \over 2 \pi \xi} \; . \label{Tphysical}
\ee
This can also be seen directly from the coordinates. The $SO(2,d)$-invariant vacuum state (analogous to the Poincar\'e-invariant vacuum in Minkowski space) is the state annihilated by the modes that have positive frequency with respect to the global time coordinate, $\tau$. Being global, $\tau$ can be assigned to each point on the entire space, (\ref{embed}), in a single-valued manner. But (\ref{RindlerObs}) then implies that the Rindler time $t$ must have an imaginary period of $2 \pi L$. Thus the Green's function of the $SO(2,d)$-invariant vacuum, when expressed in Rindler coordinates is similarly periodic in imaginary time, indicating that an Unruh detector carried by the Rindler observer will record a temperature. Finally, the proper time of the Rindler observer has an extra factor of $\sqrt{-g_{tt}}$, giving precisely (\ref{Tphysical}). Later, we will derive this temperature from the two-point correlation function in the boundary theory.

Next consider the entropy. The horizon is at $\xi = 0$. As in flat Rindler space, the area of the horizon in Rindler-AdS space is infinite:
\be
A_H ({\rm AdS}_{d+1}) \sim L^{d-2} \int_0^\infty \sinh^{d-2} (\chi/L) d \chi \; .
\ee
However, the entropy density, $s$, is finite and obeys the universal relation:
\be
s = \frac{1}{4 G_{d+1}} \; .
\ee
For three-dimensional rotating Rindler space (\ref{Rotating}), the temperature and entropy are
\be
T = \frac{1 - \beta^2}{2 \pi L} \qquad S = \f{1}{4G_3}\int (1 - \beta^2) d\chi_r \; ,
\ee
where $\beta$ is the rotation parameter, $-1 \leq \beta \leq 1$. The event horizon is still at $\xi = 0$ and the entropy is of course infinite.

\section{The Boundary Theory}

We are now interested in the holographically dual theory, which defines quantum gravity in Rindler-AdS space. As emphasized earlier, Rindler-AdS is simpler to study than eternal AdS black holes. Rindler-AdS space does not have singularities and the precise form of the boundary CFT is known in certain cases. Now, as usual in AdS/CFT \cite{Maldacena,GKP,WittenHolography}, 
in the limit of large $N$ and large 't Hooft coupling, the string partition function can be approximated at saddle point by the exponential of the classical supergravity action:
\be
Z[\phi_0(x)]_{\rm CFT} =
 \langle e^{i \int_{\partial AdS} \phi_0(x) {\cal O}(x)} \rangle \approx e^{i S_{\rm sugra}[\phi(z,x)]} \; ,
\ee
where the bulk field $\phi(z,x)$ takes the value $\phi_0(x)$ on the boundary $\partial AdS$. In the Euclidean formulation, $\phi_0(x)$ acts as a source term in the CFT, and specification of the boundary field $\phi_0(x)$ (along with the assumption of regularity in the interior) uniquely determines the bulk field, which can be determined using the bulk-boundary propagator. Thus bulk fields are dual to boundary sources. However, there are additional subtleties in the Lorentzian version of the correspondence  \cite{vjetal,vjtrivedietal} because of the existence of normalizable modes in the bulk. These are bulk excitations that do not change the leading (in $z$) contribution to the boundary value of the field, $\phi_0(x)$. The normalizable modes are dual to states in the boundary theory. For our present purpose, we will ignore the contribution of the normalizable modes and just analytically continue the bulk-boundary propagators defined in Euclidean signature in order to study the various boundary correlation functions in Lorentzian signature. We will also focus on AdS$_3$ for computational convenience; most of the results can be extended without loss of generality to higher dimensions. Below we will first recover the thermodynamics from the CFT. Then we will perform a calculation that indicates how the boundary theorist could perceive the horizon. Remarkably, the calculation indicates that at least partial information is available to the CFT about events that are across the Rindler horizon.

\subsection{Temperature and Two-Point Correlators}

We take the complete Hilbert space of conformal operators to be given by a direct product of two Hilbert spaces, ${\cal H} = {\cal H}_1 \otimes {\cal H}_2$. We also take the complete state to be an entangled state of the two CFTs, as studied in \cite{Raamsdonk,eternalAdSbh,Marcelo}:
\be
|\Psi\rangle = {1\over \sqrt{Z(\beta)}} \sum_n \!\! e^{-\beta E_n /2} |E_n\rangle_1 \times |E_n\rangle_2 \; .\label{EntangledState}
\ee
This state corresponds to the vacuum of the boundary theory in global AdS spacetime. Also, the temperature ($1/\beta$) of this entangled state is unique and related to the AdS scale, as we will see later. All expectation values of the conformal operators are taken with respect to the entangled state given by (\ref{EntangledState}). In order to compute correlation functions in the boundary theory, one needs the explicit form of the bulk-boundary propagator $ K(\xi,\chi,t;\chi_0,t_0)$ defined by
\be
 \phi(\xi,\chi,t) = \int \!\! K(\xi,\chi,t;\chi_0,t_0)\phi_0(\chi_0,t_0)\!~ d\chi_0 dt_0 \; .
\ee
Here the point $(\chi_0,t_0)$ acts as a source on the boundary while the bulk point $ (\xi,\chi,t)$ is the sink. In AdS$_3$, the bulk-boundary propagator for a minimally coupled massive scalar field, upto normalization, is
\be
 K(\xi,\chi,t;\chi_0,t_0) = {1\over {\left[\sqrt{1 + {\xi^2\over L^2}}\cosh({{\chi - \chi_0}\over L}) - {\xi\over L}\cosh({{t-t_0}\over L}) \right]^\Delta}} \; . \label{BulkBoundaryPropagator}
\ee
Here $\Delta = 1 + \sqrt{1 + m^2}$ is the conformal dimension of the boundary operator dual to a bulk scalar of mass $m$. The bulk-boundary propagator satisfies the massive wave equation in Rindler-AdS coordinates and is valid as long as both the source and sink happen to be on the same side of the Rindler horizon i.e. when the conformal operaters are inserted on the same boundary. As $ \xi \rightarrow \infty$, $K$ becomes a delta function supported at $ \chi = \chi_0$ and $t=t_0 $. Using the standard rules for AdS/CFT \cite{GKP,WittenHolography}, the two-point function between conformal operators inserted on the same boundary is 
\be 
 \langle  {\cal O} (\chi_1,t_1)  {\cal O} (\chi_2,t_2) \rangle = {1\over {\left[\cosh({{\chi_1 - \chi_2}\over L}) - \cosh({{t_1-t_2}\over L}) \right]^{1 + \sqrt{1 + m^2}}}} \; . \label{2pointFn}
\ee
The two-point functions has a periodicity of $2 \pi L$ in imaginary time; evidently the boundary CFT is thermal in nature, as mentioned previously for the entangled state (\ref{EntangledState}), with $\beta = 2 \pi L$. This is in agreement with the fact that the temperature of the Rindler horizon is indeed $T_H = {1\over 2 \pi L}$. Hence the boundary theory gives the correct horizon temperature.

To evaluate the bulk-boundary propagator when the sink is on the other side of the horizon, we analytically continue the time as $t \rightarrow t - i \pi L$, as can be seen from (\ref{Rindlercoords}). The bulk-boundary propagator then becomes
\be
 K(\xi,\chi,t;\chi_0,t_0) = {1\over {\left[\sqrt{1 + {\xi^2\over L^2}}\cosh({{\chi - \chi_0}\over L}) + {\xi\over L}\cosh({{t-t_0}\over L}) \right]^{1 + \sqrt{1 + m^2}}}} \; .
\ee
Using the above bulk-boundary propagator and the rules of AdS/CFT we arrive at the two-point function of operators inserted on the opposite boundaries
\be 
 \langle  {\cal O}_1 (\chi_1,t_1)  {\cal O}_2 (\chi_2,t_2) \rangle = {1\over {\left[\cosh({{\chi_1 - \chi_2}\over L}) + \cosh({{t_1-t_2}\over L}) \right]^{1 + \sqrt{1 + m^2}}}} \; . \label{Spacelike2pointFn}
\ee
The two-point function is nonsingular because the operators are always spacelike separated. The reason the expectation value does not vanish even though the operators on opposite boundaries commute is that the CFTs are entangled. 

In general, correlation functions can be calculated in global AdS coordinates and then transformed to Rindler-AdS coordinates. This is of course no different from what happens in flat Rindler space for which (bulk) correlation functions can be calculated in standard Minkowski coordinates and then transformed to Rindler coordinates.

\subsection{Entropy}

First consider the entropy in higher dimensions. Specializing to AdS$_5$, the Rindler horizon has entropy 
\be
S_{\rm Rindler} = \lim_{\chi_0 \rightarrow \infty}{\pi L^2 \over G_5} \int_0^{\chi_0} \! \! {\sinh^2 (\chi /L) d\chi} \; , \label{Rindlerentropy}
\ee
which diverges as expected. The coordinate $\xi$ scales the boundary theory. Specifically, for $AdS_5$, the dual theory is ${\cal N} = 4$ SYM theory, with a gauge field, four Weyl spinors and six conformally coupled scalars, all in the adjoint of $SU(N)$. The number of degrees of freedom is thus $15 N^2$. The size of the gauge group is related to the AdS radius by
\be
N^2 =  {\pi L^3 \over 2 G_5} \; . \label{gaugegroup}
\ee
A priori, there are now two ways of calculating the entropy from the dual theory: as the entropy of a gas of thermal free fields, and as entanglement entropy. The free field entropy computation for a thermal CFT is done using the standard result
\be 
S_{CFT} = {2\over 3} \pi^2 N^2 V_{CFT}T_{CFT}^3 \; . \label{Freefieldentropy}
\ee
Evaluating this ``holographically'' implies substituting boundary data into the above expression. At fixed $\xi = \xi_0 \gg L $, the boundary metric is
\be
ds^2 = \xi_0^2 \left [-{dt^2 \over L^2} + {d\chi^2 \over L^2} + \sinh^2 \left ({\chi \over L} \right ) d\Omega_2^2 \right] \; . \label{boundarymetric}
\ee
The horizon temperature is given by $T_H = {1 \over 2 \pi L} $ and the physical temperature at the boundary is
\be
T_{CFT} = {T_H \over \sqrt{-g_{tt}}} = {1 \over {2 \pi \xi_0}} \; , \label{CFTTemp}
\ee
and $V_{CFT}$ is given by
\be
V_{CFT}= \lim_{\chi_0 \rightarrow \infty}{4 \pi \xi_0^3 \over L}\!\! \int_0^{\chi_0} \! \! {\sinh^2 (\chi /L) d\chi}  \; . \label{Volume} 
\ee
Using (\ref{gaugegroup}), (\ref{CFTTemp}), (\ref{Volume}) and inserting them into (\ref{Freefieldentropy}), we see that the free field CFT entropy scales in the same manner as (\ref{Rindlerentropy}), albeit with 
\be
S_{CFT} = {1\over 6} S_{\rm Rindler} \; .
\ee
This familiar numerical disagreement is presumably because of the fact that we have assumed the large $N$ limit and large 't Hooft coupling. In this approximation, the entropy of the boundary theory is computed using the results for a free field CFT. In the exact case however, the CFT could be a fully interacting field theory; we do not yet understand how to calculate the entropy for such a theory directly.

So far this is all mostly familiar. We can do much better for Rindler-AdS$_3$.
For (\ref{3dimRAdS}), the Bekenstein-Hawking entropy is given by
\be
S_{BH} = \frac{A}{4G_3}=\frac{\int d\chi}{4G_3} \; .
\ee
The Euclideanized boundary metric for (\ref{3dimRAdS}) is given by
\be
ds_{\rm boundary}^2 = d\tau^2 + d\chi^2 \; , \label{Euclideanboundary}
\ee
where $\tau \sim \tau + \beta = \tau + 2\pi L$, and the last equality follows from the fact that the boundary two-point function (\ref{2pointFn}) is periodic in imaginary time with period $\beta = 2 \pi L$. 
Since by the AdS/CFT correspondence $Z_{AdS} = Z_{CFT}$, we can now use the Cardy formula to calculate the entropy of the CFT:
\be
S_{CFT} = \frac{\pi}{3 \beta} c ~ {\rm Volume}= \frac{\pi}{3} \frac{3L}{2 G_3} \frac{1}{2 \pi L} \int d\chi = S_{BH} \; , \label{NR-Entropy} 
\ee
where $c = \frac{3L}{2 G_3}$ is the central charge of the unitary CFT as calculated by Brown and Henneaux \cite{BrownHenneaux}. Of course the entropy of the Rindler horizon is infinite, but it is very interesting that the entropy densities are now in precise agreement.

We can also use the Cardy formula for the rotating CFT:
\be
S_{CFT} = \frac{\pi}{3} c T ~ {\rm Volume}= \frac{\pi}{3} \frac{3L}{2 G_3} \frac{1- \beta^2}{2 \pi L}\int d\chi_r = S_{BH} \; . \label{R-Entropy} 
\ee
Once again the CFT entropy density and the Bekenstein-Hawking entropy density are in precise agreement, including the numerical factor. Under the diffeomorphism (\ref{NR-to-R}), the volume element transforms as $d\chi \rightarrow (1 - \beta^2)d\chi$, and therefore (\ref{NR-Entropy}) and (\ref{R-Entropy}) both have the universal entropy density $1/4G$. 

\subsection{Relation between Rindler-AdS space and AdS black holes}

Let us pause here to comment briefly on the relation between Rindler-AdS space and black holes in anti-de Sitter space. From the outset, it is important to clarify that Rindler-AdS space is not the near-horizon limit of black holes in AdS; the near-horizon limit of all non-extremal black holes, including black holes in AdS space, is flat Rindler space.

The existence of an ergosphere in rotating Rindler-AdS space recalls the rotating BTZ black hole. Indeed, rotating Rindler-AdS space is related to the rotating BTZ black hole \cite{BTZ,BTZH} via 
\be
\chi_r \sim \chi_r + 2\pi  \; .
\ee
A change of coordinates
\be
\xi = \sqrt{\frac{r ^2 - 1}{1 - \beta ^2}}
\ee
puts the metric in the familiar BTZ form:
\be
ds^2 = -\frac{(r ^2 - 1)(r ^2 - \beta ^2)}{r^2}dt_r^2 + \frac{r^2}{(r ^2 - 1)(r ^2 - \beta ^2)}dr^2 + r^2 \left (d\chi_r - \frac{\beta}{r^2}dt_r \right )^{\! 2} \; .
\ee
Rindler-AdS is thus the universal cover for the BTZ black hole \cite{Emparan,Myers,Lowe,Lowe2,Vanzo}. The black hole solution is obtained by making an identification in a direction perpendicuar to $\partial_t$ at the boundary. However, there is an important difference between Rindler-AdS space and the BTZ black hole. The identification breaks the symmetry group down from $SL(2,R) \times SL(2,R)$ to $SL(2,R) \times U(1)$. Consequently, the freedom of picking out the time direction is lost; neither the event horizon nor the ergosphere of the BTZ black hole is observer-dependent. Put another way, the identification $\chi_r \sim \chi_r + 2 \pi $ gives the two-dimensional boundary Minkowski space a cylinder topology. But special relativity on a cylinder has a preferred frame, singled out by the identification \cite{BarrowLevin,Greene}. Hence there is a preferred direction of time.

That Rindler-AdS$_3$ is the universal cover of the BTZ black hole also means that two-point functions in the CFT for BTZ black holes are infinite sums of Rindler-AdS two-point functions summed over all image points. For example, for operators inserted on opposite boundaries, the BTZ two-point correlator is 
\begin{eqnarray}
 \langle  {\cal O}_1 (\chi_1,t_1)  {\cal O}_2 (\chi_2,t_2) \rangle_{\rm BTZ} & \sim & \sum_{n = -\infty}^{n= +\infty} {1\over {\left[\cosh({{\chi_1 - \chi_2 + 2 \pi n}\over L}) + \cosh({{t_1-t_2}\over L}) \right]^{1 + \sqrt{1 + m^2}}}} \nonumber \\ 
& \sim  & \sum_{n = -\infty}^{n= +\infty}  \langle  {\cal O}_1 (\chi_1 + 2 \pi n,t_1)  {\cal O}_2 (\chi_2,t_2) \rangle_{\rm Rindler} \; .
\end{eqnarray}
The relative simplicity of the two-point function in Rindler-AdS is, as we shall see below, another one of the advantages of Rindler-AdS as a model spacetime in the study of horizons.

\subsection{The Omniscient CFT}

It is now widely believed, if not proven, that the process of black hole formation and evaporation is unitary. The existence of a unitary conformal field theory dual to anti-de Sitter space lends support to this belief, as the formation and evaporation of AdS black holes is presumably a process that has a dual description within a unitary theory. Nevertheless, a detailed account of how information emerges from a black hole is far from clear. Here we will take a step in that direction by showing that the dual CFT can tell whether an infalling source has crossed the horizon. In fact, the CFT even has partial information about events that happen across the horizon. This is promising because, from the local bulk point of view, the horizon is a nondescript place; by contrast, gauge/gravity duality is nonlocal and it is precisely in a theory with nonlocality that one expects to be able to evade the paradoxes of black holes.

There are of course several different ways to probe the horizon. Here we will consider ``switching on'' a point source which freely falls into the Rindler horizon, before being ``switched off'' after the passage of some finite interval of proper time. The source couples to a bulk field which, for simplicity, we will take to be a free scalar field. The boundary value of the bulk field in turn plays the role of a coupling constant in the boundary CFT. Consider, as an analogy, a Reissner-Nordstrom black hole. The bulk field here would be the electromagnetic field and a source would be any charge or current configuration. For the purpose of understanding information retrieval, one might like to send in a source that carries no coarse-grained hair (i.e. no mass, charge, or angular momentum) such as, say, an electric dipole, to test whether the CFT can determine what was thrown in. The alternative to throwing in a source would be to send in some excitation of the field itself; this would be analogous to probing our Reissner-Nordstrom black hole by sending in an electromagnetic wave which will propagate on null trajectories. Hence in light-cone or Eddington-type coordinates, the wave would have a constant ingoing null coordinate and we would not be able to distinguish the moment the packet crossed the horizon from any earlier moment. The advantage of sending in a source is that it can travel on a timelike trajectory, for which the ingoing null coordinate time varies along the trajectory. Therefore, by considering the signatures of the ``switching on'' and ``switching off'' processes of our infalling source, we will see that the CFT can tell whether the source is switched on or off even after it crosses the Rindler horizon.

Our goal then is to study the response of the boundary operator that is dual to the bulk field, as the source falls into the horizon. For simplicity we consider a point-like source, but this is a good approximation since even more a realistic source would have its wave-packet blue-shifted and increasingly localized as it approaches the horizon. In principle, we could also explicitly construct a CFT operator dual to the infalling source. Such a construction would depend on the nature of the source. For example, consider the case where the bulk field is the metric. Then the infalling source would be described by the bulk matter energy-momentum tensor, which in turn would be made up of bulk matter fields. These couple to the CFT through their boundary values; there is a considerable literature on how to create localized bulk fields (that constitute the infalling source) through smearing functions at the boundary \cite{Lowe, Lowe2, Raju}. Therefore, in principle there is no problem in describing the infalling bulk source in the CFT. However, such a construction is somewhat irrelevant to the question we are trying to consider here i.e. how do the boundary one-point functions of operators dual to the bulk field (that is sourced by the infalling source) yield information on across-horizon physics?
 
In order to describe an infalling source, we need to define the Rindler coordinates beyond the horizon i.e. into the region $(X^1)^2 - (X^0)^2 < 0$. To that end, we transition to ingoing Eddington-Finkelstein (EF) coordinates by defining 
\begin{eqnarray}
r &\equiv& \frac{\xi^2}{2L} \nonumber \\
v &\equiv& t + \int \frac{dr}{\frac{2r}{L} \sqrt{1 + \frac{2r}{L}}} = t + \f{L}{2}\ln \left[\f{\sqrt{1 + \f{2 r}{L}} - 1}{\sqrt{1 + \f{2 r}{L}} + 1} \right] \; . \label{Eddingtontime}
\end{eqnarray}
With this, the Rindler-AdS$_3$ metric in EF coordinates becomes
\be
ds^2 = -\frac{2r}{L}dv^2 + \frac{2dvdr}{\sqrt{\left(1 + \frac{2r}{L} \right)}} + \left(1 + \frac{2r}{L}\right)d\chi^2 \; . \label{EddingtonMetric}
\ee
The ranges of the coordinates is $-L/2 < r < \infty$ and $-\infty < v < \infty$, with the region outside the horizon being $0 < r < \infty$. In particular, these coordinates are perfectly smooth at the future horizon $r = 0$. These coordinates span one patch of the Rindler-AdS space time ($-\frac{L}{2} < r < \infty$). In the Penrose diagram, the entire space time can be viewed as an infinite concatenation of such identical patches, in the direction of the global time coordinate. The boundary metric at large $r$ is
\be
ds^2_b = \frac{2r}{L} \left(-dv^2 + d\chi^2 \right) \; , \label{EddingtonBoundaryMetric}
\ee
which is conformally flat, an advantage of working in three dimensions.

In order to describe a source falling into the Rindler horizon, we consider timelike radially ingoing geodesics in Rindler-AdS$_3$. Since the metric is invariant under translations of the $v$ coordinate, the momentum component $p_v$ is conserved along geodesics. Since $p_v = m u_v$ (where $u^a$ is the velocity vector), and setting $m \equiv 1$, we have that $u_v$ is conserved. For simplicity, let the conserved value of $u_v$ be $-1$. Then setting $\chi$=const so that $u^\chi = 0$ (which corresponds to radial infall) we have
\be
(u^r)^2 + \f{4r^2}{L^2} = 1 \; . \label{r-vel}
\ee
Choosing the initial condition  $r(0) = L/2$ and using $u_v = -1$, we 
find that the source's geodesic trajectory is given by
\begin{eqnarray}
r_J(\tau) &=& \frac{L}{2}\cos \left(\frac{2\tau}{L} \right) \nonumber \\
v_J(\tau) &=& \f{L}{2} \ln \left[\f{1 + \sin \left(\f{2\tau}{L} \right)}{(\sqrt{2}\cos\left(\f{\tau}{L} \right) + 1)^2} \right] \; , \label{EOM}
\end{eqnarray}
where $\tau$ is the proper time.

The conditions are chosen such that, at $\tau = 0$, we have $r = L/2$ and $v = - L \ln  (1+ \sqrt{2})$. 
The source exits the patch covered by Eddington coordinates at $\tau_{\rm max} = L \pi/2$ for which $v_{\rm max} = 0$. In particular, the source crosses the Rindler horizon at
\be
\tau_h = L \frac{\pi}{4} \quad , \quad r_h = 0  \quad , \quad v_h = - \frac{L}{2} \ln 2 \; .
\ee

We now consider a bulk scalar field, $\phi$, sourced by a freely falling localized source, $J$, which we model as
\be
J=\int_{\tau_i}^{\tau_f} d\tau~\delta (r-r_J(\tau)) \delta (v-v_J(\tau)) \delta (\chi-\chi_J(\tau)) \; , \label{source}
\ee
where $r_J(\tau)$ and $v_J(\tau)$ are given by (\ref{EOM}), and $\chi_J(\tau) = 0$ for simplicity. In addition, we require the source to get ``switched on'' at a certain instant with proper time $\tau_i \geq 0$, then traverse the geodesic path (\ref{EOM}) before getting ``switched off'' or terminated at a later proper time, $\tau_f$.

In order to describe the infall of the source into the horizon from the boundary perspective, we use the basic AdS/CFT tool
\be
\int_{\rm bulk} {\cal D} \phi \, e^{iI[\phi]} = \left \langle e^{\int \!  \phi_0 {\cal O}} \right \rangle_{CFT} \; , \label{AdS-CFT}
\ee 
where $\phi_0$ is the boundary value of the bulk field $\phi$. The AdS/CFT dictionary mandates the equivalence of the bulk and boundary vacua. We evaluate correlation functions with respect to the state (\ref{EntangledState}), which is the AdS analog of the Hartle-Hawking vacuum state. Using the SUGRA approximation, we can approximate the bulk path integral by its saddle-point
\be
\int_{\rm bulk} {\cal D} \phi \, e^{iI[\phi]} \sim e^{i I [\phi_{\rm cl}]} \; , \label{sugra}
\ee 
where $I [\phi_{\rm cl}]$ is the action for the classical field configuration. In order to evaluate the bulk action, we need to first find $\phi_{\rm cl}$. Given $J$, we can solve for the bulk scalar field as 
\be
\phi_{\rm cl}(r,\chi,v) = \int G(r,\chi,v;r',\chi',v')J(r',\chi',v')dr' d\chi' dv' \; ,
\ee
where $G(r,\chi,v;r',\chi',v')$ is the bulk-bulk propagator. For our source (\ref{source}) we have
\be
\phi_{\rm cl}(r,\chi,v) = \int_{\tau_i}^{\tau_f} G(r,\chi,v;r_J(\tau),\chi_J(\tau),v_J(\tau))d\tau \; . \label{bulkfield}
\ee
An important point to note is that the propagators that arise in path integrals, such as on the left-hand side of (\ref{AdS-CFT}), are Feynman propagators; Feynman's $i\epsilon$ prescription is necessary for path integrals to converge. Hence we must use the Feynman propagator to evaluate $\phi_{\rm cl}$ in order to be consistent with our setup. This is very important since the Feynman propagator, which crucially does not vanish at spacelike separation, can yield signatures about across-horizon physics. 

The boundary value, $\phi_0(\chi, v)$, of the scalar field can be obtained by taking $\displaystyle \lim_{r\rightarrow \infty} \phi_{\rm cl}(r,\chi,v) = \phi_0 (\chi,v)$.  The explicit form for the bulk-bulk Feynman propagator for AdS$_3$ was derived in \cite{Danielsson} and is given by
\begin{equation}
G(r_1,\chi_1,v_1;r_2,\chi_2,v_2) \sim \gamma^{\Delta}~_2F_1 \left (\frac{\Delta}{2}, \frac{\Delta}{2} + \frac{1}{2},\Delta,\gamma^2 \right ) \; , \label{bulk-bulk}
\end{equation}
where $\Delta = 1 + \sqrt{1 + m^2}$. The bulk-bulk propagator (\ref{bulk-bulk}) is calculated using normalizable modes in Poincar\'e coordinates; the Poincar\'e vacuum is equivalent to the global vacuum \cite{Danielsson}. Here $\gamma$ is related to the AdS invariant geodesic distance,
\be
\gamma = \f{L^2}{X_{1}^a X_{2}^b \eta_{ab}} \; ,
\ee
for any two vectors $X^a_1$ and $X^a_2$, where $\eta_{ab}$ is the Minkowski metric in the embedding space (i.e. with two time directions).
In EF coordinates (see appendix), we find that
\be
\gamma = \frac{L^2}{ + \sqrt{+ \frac{4r_1 r_2}{L^2}}\cosh\left( \frac{v_2 - v_1 - f(r_2) + f(r_1)}{L} \right) -\sqrt{\left(1+ \frac{2r_1}{L} \right) \left(1+ \frac{2r_2}{L} \right)}\cosh\left( \frac{\chi_2 - \chi_1}{L} \right)} \; .\label{gamma}
\ee

According to the AdS/CFT correspondence, at large $N$ and large 't Hooft coupling, the one-point function is given by
\begin{equation}
\langle {\cal O}(v,\chi)\rangle = \lim_{r \rightarrow \infty} \frac{1}{\sqrt{-h}}\frac{\delta I}{\delta \phi_0 (v,\chi)} \; , \label{One-point}
\end{equation}
Here $h$ is the determinant for the boundary metric (\ref{boundarymetric}). Let us first evaluate the action. The action for the field $\phi$ is
\be
I[\phi] = \int \left ( -\frac{1}{2} \left(\partial \phi \right)^2 - \frac{1}{2}m^2 \phi^2 + J \phi \right ) d\chi dv dr \; .\label{action}
\ee 

Integrating (\ref{action}) by parts, and separating the bulk and the surface terms, we get for the variation of the action
\be
\delta I[\phi_{\rm cl}] \sim \int g^{\mu \nu} \delta \phi_{\rm cl} \, \partial_{\mu}\phi_{\rm cl} \, d\Sigma_\nu \; ,
\ee
where $d\Sigma_\nu$ is the surface normal to the $v$ coordinate and the variation of the bulk term vanishes on-shell. Since we wish to evauate this action at the boundary, i.e. at $r\rightarrow \infty$, using the above expression and (\ref{One-point}), the one-point function is
\be
\langle {\cal O}\rangle \sim \lim_{r \rightarrow \infty} \frac{\sqrt{-g}}{\sqrt{-h}}g^{r\mu}\partial_{\mu}\phi_{\rm cl} \; ,\label{one-point-expression}
\ee 
as one power of $\phi$ is pulled down by differentiation. We now plug in (\ref{bulkfield}) to get 
\be
\langle {\cal O}\rangle \sim \lim_{r \rightarrow \infty} \frac{\sqrt{-g}}{\sqrt{-h}}g^{r\mu}\partial_{\mu}\int_{\tau_i}^{\tau_f} G(r,\chi,v;r_J(\tau),\chi_J(\tau),v_J(\tau))d\tau \; . \label{propagator.derivative}
\ee
Finally, we assume a massless scalar field $m=0 \Rightarrow \Delta=2$ for ease of calculation, $\chi_J = 0$, and insert (\ref{EOM}), (\ref{bulk-bulk}), and (\ref{gamma}) into the above expression. Next, we notice from (\ref{gamma}) that $\gamma$ goes to zero as $r \to \infty$. We can therefore perform a power series expansion of the hypergeometric function for small $\gamma$ in terms of  Pochhammer symbols. We then get
\begin{eqnarray}
\lim_{r \rightarrow \infty} \pa_r \left[\gamma^2~_2F_1 \left (1, 3/2,2,\gamma^2 \right )\right] &=& \lim_{r \rightarrow \infty} \frac{\pa}{\pa \gamma^2} \left[\gamma^2 \left(1 + \frac{3 \gamma^2}{4} + ... \right)\right] \frac{\pa \gamma^2}{\pa r} \nonumber \\
&=& \lim_{r \rightarrow \infty} \left[1 + \frac{3 \gamma^2}{2} + ...\right] \frac{\pa \gamma^2}{\pa r} \; ,\label{intermediate.steps}
\end{eqnarray}
where, from (\ref{gamma}), we have 
\be
 \lim_{r \rightarrow \infty} \frac{\pa \gamma^2}{\pa r} = \frac{-1}{r^2_{\infty} \left[\sqrt{1+ \cos\left(\frac{2\tau}{L} \right)} \cosh (\chi) - \sqrt{\cos\left(\frac{2\tau}{L} \right)}\cosh \left(\frac{v}{L} - g(\tau) \right)  \right]^{2}} \; .
\ee
Here $r_\infty$ is the infrared cutoff that marks the surface on which the CFT lives. Therefore in the large $r=r_\infty$ limit, only the first term in (\ref{intermediate.steps}) contributes. Noting that in the large $r$ limit, $\sqrt{-h} \rightarrow \frac{2r_\infty}{L}~,~~g^{rr} \rightarrow \frac{4r^2_\infty}{L^2} $, we have for the one-point function
\be
\langle {\cal O}(v,\chi)\rangle \sim \int_{\tau_i}^{\tau_f} \frac{d\tau}{r_\infty \left[\sqrt{1+ \cos\left(\frac{2\tau}{L} \right)} \cosh (\chi) - \sqrt{\cos\left(\frac{2\tau}{L} \right)}\cosh \left(\frac{v}{L} - g(\tau) \right)  \right]^{2}} \; , 
\ee
where $g(\tau) = v_J (\tau) - \f{L}{2}\ln \left[\f{\sqrt{1 + \f{2r_J (\tau)}{L}} - 1}{\sqrt{1 + \f{2r_J (\tau)}{L}} + 1} \right] = \f{1}{2} \ln \left[\f{1 + \sin \f{2\tau}{L}}{\cos \f{2\tau}{L}} \right]$. The appearance of the $\frac{1}{r_\infty}$ factor is consistent with the scaling dimensions of the operator ${\cal O}$. The above integral can be further simplified to yield
\be 
\langle {\cal O}(v,\chi)\rangle \sim \int_{\tau_i}^{\tau_f} \! \! \! \frac{4~ d\tau}{r_\infty \left[2 \sqrt{1 + \cos \left(2 \tau/L \right)} \cosh \chi - e^{-v/L} \sqrt{1 + \sin \left(2 \tau/L \right)} - e^{v/L} \frac{\cos \left(2 \tau/L \right)}{\sqrt{1 + \sin \left(2 \tau/L \right)}} \right]^2} \; . \label{Response}
\ee  

\subsection{Signatures of Across-Horizon Physics}

First, let us consider the one-point function when the source is both switched on and switched off outside the horizon. For instance, we could take $\tau_i = 0$ and $\tau_f = L\pi/6 < \tau_h$. Setting $\chi =0$ and performing the integral (\ref{Response}), we obtain 
\be
\langle {\cal O}(v,0)\rangle \sim \frac{1}{r_\infty \left(\sqrt{2} - \cosh(v/L) \right) \left(\sqrt{6} - \sqrt{3}\cosh(v/L) + \sinh(v/L) \right)} \; . \label{Response.outside}
\ee
Notice that the one-point function has {\em four} poles at
\begin{eqnarray}
u_i = L \ln (\sqrt{2} + 1) & , & v_i =L \ln (\sqrt{2} - 1) \nonumber \\
u_f = \frac{L}{2} \ln (2 + \sqrt{3})(5 + 2\sqrt{6}) & , & v_f = \frac{L}{2} \ln (2 + \sqrt{3})(5 - 2\sqrt{6})
  \; .\label{poles.outside}
\end{eqnarray} 
Here $u$ and $v$ are ingoing and outgoing Eddington-Finkelstein coordinates; $u$ is related to the $v$-coordinate by $u = v - 2f ( r)$, where $f (r )$ is given by the log term in (\ref{Eddingtontime}). We have expressed two of the poles in terms of $u$ coordinates for reasons that will be clear soon.

Now, consider the case where the source switches off only after it crosses the horizon. For example, choose $\tau_i = 0$ and $\tau_f = L \pi/2 > \tau_h$. Evaluating the integral, we find 
\be
\langle {\cal O}(v,0)\rangle \sim \frac{1}{r_\infty \left(\sqrt{2} - \cosh(v/L) \right) \sinh (v/L)} \; .\label{Response.across}
\ee
In this case the one-point function has only {\em three} poles. They are at
\begin{eqnarray}
u_i = L \ln (\sqrt{2} + 1) & , & v_i = L \ln (\sqrt{2} - 1) \nonumber \\
v_f & = & 0 \; . \label{poles.across}
\end{eqnarray} 

\begin{figure}[hbtp]
 \centering
  \epsfysize=6 cm
  \epsfbox{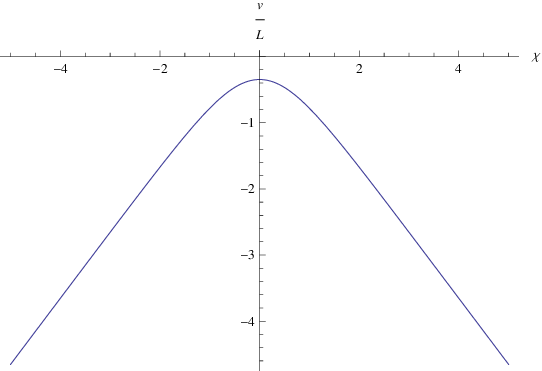}
 \centering
  \caption{The locus of points on the boundary where there are poles coming from one endpoint of the source trajectory. The specific values plotted are for the case where the source switches off precisely on the horizon, for which there are only $v$ poles coming from the intersection of the past light cone of the endpoint with the hypersurface on which the CFT lives.}
\label{fig:chipoles}
 \end{figure}

The appearance of poles in the one-point function is easy to understand. We considered an idealized source which is nonzero only for a finite interval of proper time, $\tau_i \leq \tau \leq \tau_f$. As a result, the field $\phi_{\rm cl}$ is discontinuous at the endpoints ($\tau_i,\tau_f$) since at these points we abruptly switch the source on and off. But the one-point function is related to the derivative of the field (\ref{propagator.derivative}). The poles therefore come from taking the derivative of a discontinuous field. The discontinuity in the field propagates towards the AdS boundary along light-like trajectories.
Moreover, since we are using the Feynman propagator, the propagation of these signals occur via the retarded (the $u$ poles) as well as the advanced component (the $v$ poles) of the propagator. In a certain sense, these poles indicate the creation and annihilation of the source from a boundary theory perspective. There are also poles in the $\chi$ (spatial) direction on the boundary. That is because the locus of poles is the intersection of the constant $r$ hypersurface where the CFT lives with the past/future light cone emanating from the endpoint. See Figure \ref{fig:chipoles}.

\begin{figure}[hbtp]
  \epsfysize=6.0 cm
  \epsfbox{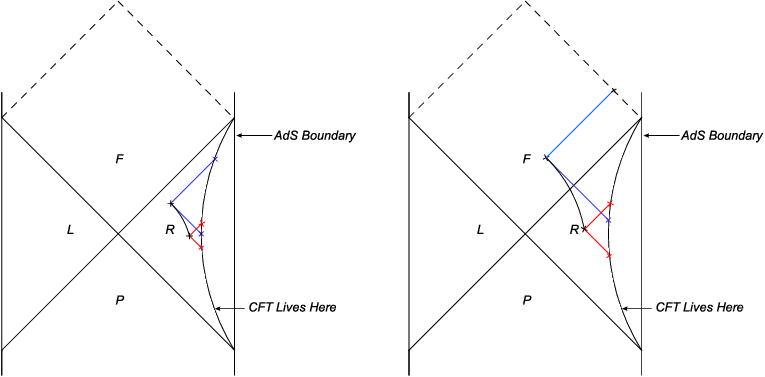}
\caption{a) The left figure illustrates when the source is active for a certain time period outside the horizon in the right Rindler wedge (R). The red and blue lines indicate signals propagating towards the AdS boundary which correspond to the creation and annihilation of the source respectively. The four poles are indicated on the boundary where the CFT lives. b) The right figure shows a source that crosses the horizon. It is evident that the retarded signal from the annihilation (or switching off) of the source no longer reaches the CFT boundary, and therefore the CFT perceives just three poles as shown. The dashed lines indicate the boundary of the Eddington-Finkelstein coordinates.}
\label{fig:uvpoles}
 \end{figure}

Now the crucial point is that, once the source crosses the horizon, there is no pole corresponding to the outgoing Eddington coordinate $u$ when the source switches off at $\tau_f$. This is because once past the horizon, retarded signals from the source do not reach the surface where the CFT lives. This is schematically illustrated in Figure \ref{fig:uvpoles}.

Evidently, the poles of the one-point function, $\langle {\cal O} \rangle$, allows the boundary theorist to determine whether the source was annihilated before or after crossing the horizon. If there are four poles, the source switched off before it reached the Rindler horizon; if there are only three poles, it means that the source switched off after horizon-crossing. But in order to determine whether the source switches off before or after the horizon, the boundary theorist has to observe the one-point function for all time. For example, a source that is switched off just infinitesimally before crossing the horizon will contribute a future-light-cone ($u$) pole in the near-infinite future. So the boundary theorist has to wait till future infinity to determine whether there are three poles or four.

In fact, the boundary theorist even acquires partial information about the  location of the switching off event, even if that event was across the horizon. In our radial infall scenario, we have effectively suppressed the $\chi$ coordinate and the location of a switching on/off event is characterized by its $u$ and $v$ coordinates. If the source switches off before it traverses the horizon, the CFT pole structure records both the $u$ and the $v$ values of the event so that its precise location can be identified. Even if the source switches off after it crosses the horizon, the CFT still knows about the $v$ value of the event. So partial information is obtained even about events that happen across the event horizon. Contrast this with a bulk observer who does not see anything fall into the horizon in finite time, and therefore would also not see any information come out in finite time. The key difference is that the boundary theorist has access to the one-point function of a CFT whose relation to the bulk is non-local -- and which can therefore encode information about events beyond the horizon.

In order to exactly read off $v$, the boundary theorist has to make certain assumptions about the geometry behind the Rindler horizon. This is implicit in our set-up since we have assumed that the geometry is pure AdS everywhere (including in the other Rindler wedge). Nevertheless, even if there were deviations from pure AdS behind the horizon, the qualitative result regarding the number of poles ``seen'' by the boundary theorist would still hold. That is, irrespective of the geometry behind the horizon, the boundary theorist would still perceive only three poles in the one-point function if the source were to get switched off behind the horizon. Only the precise location of the third pole (i.e. the ``$v$'' coordinate of the pole in the boundary theory) would be sensitive to beyond-horizon metric perturbations. 

We have used a boundary correlation function to detect a simple signature of the across-horizon physics of an infalling bulk source. Notably, the boundary correlation function is accessible to a boundary theorist with access to only one CFT. However, to actually create the bulk source in the boundary theory \cite{Raju}, one needs both the right as well as the left CFTs. Hence a boundary theorist with access to only one CFT cannot single-handedly set up the experiment. Nevertheless, a one-sided boundary theorist can read off the results of the experiment -- there are distinct and measurable effects for the one-sided CFT depending on whether the source switches off before or after crossing the horizon -- even if such a theorist may not recognize it as an infalling source. This is precisely the spirit of our calculation. More precisely, from our construction (see Figure \ref{fig:uvpoles}) we can see that the past light cone of a switching off event in the upper Rindler wedge (F) also intersects the antipodal CFT (associated with a hypersurface in region (L)). The missing fourth pole is actually in the antipodal CFT; complete knowledge of the pole structure of both CFTs is therefore necessary to fully reconstruct switching off events in the upper Rindler wedge.

\section{De Sitter space as the boundary of Rindler-AdS}

In this section, we touch upon an alternate formulation of Rindler-AdS with a potentially wide spectrum of applications. Consider again a Rindler observer in $d+2$-dimensional Minkowski space (with two time directions) uniformly accelerating in the $X^1$ direction:
\be
X^0 = {\tilde r} \sinh (t/ L) \qquad X^1 = {\tilde r} \cosh (t/ L) \; . \label{Rindlerobservers}
\ee
This turns the flat space line element into
\be
ds^2 = - {\lf \tilde r \over
L\rt}^{\! 2} dt^2 + d{\tilde r}^2 + dX_2^2 + ... + dX_d^2 - dX_{d+1}^2 \; ,
\ee
which, indeed, is Rindler space (albeit with two time directions). Rindler observers at constant ${\tilde r}$ have proper acceleration $1/{\tilde r}$. We foliate AdS as
\begin{eqnarray}
X^0 & = & R \cos \chi \sinh (t/ L) \nonumber \\
X^1 & = & R \cos \chi \cosh (t/ L)  \nonumber \\
X^2 & = & R \sin \chi \cos \theta_1  \nonumber \\
& ... &  \nonumber \\
X^{d-2} & = & R \sin \chi \sin \theta_1 ... \sin \theta_{d-3} \cos \theta_{d-2} \nonumber \\
X^{d-1} & = & R \sin \chi \sin \theta_1 ... \sin \theta_{d-2} \cos \phi \nonumber \\
X^d & = & R \sin \chi \sin \theta_1 ... \sin \theta_{d-2} \sin \phi \nonumber \\
X^{d+1} & = & \sqrt{L^2 + R^2}  \; . \label{AlternateRindlercoords}
\end{eqnarray}
This satisfies the AdS embedding equation (\ref{embed}). The first two coordinates are of the form (\ref{Rindlerobservers}) with what we called ${\tilde r}$ now being $R \cos \chi$. Defining $r = L \sin \chi$, we finally obtain
\be
ds^2 = {dR^2 \over 1 + (R/L)^2} + \lf R/ L \rt^2 \left [ - (1 - (r/ L)^2 ) dt^2 + {dr^2 \over 1 - (r/ L)^2} + r^2  d \Omega_{d-2}^2 \right ]  \; . \label{RindlerAdSmetric-DeSitter}
\ee
We see that Rindler-AdS can also be foliated in slices that are conformal to static de Sitter space with de Sitter radius $L$ \cite{hms,daszelnikov}. The ranges of the coordinates are
\be
0 \leq R  \qquad -\infty < t < \infty \qquad 0 \leq r <
L \qquad 0 \leq \theta_i \leq \pi \qquad 0 \leq \phi < 2 \pi \; .
\ee
The coordinate $r$ is related to the polar angle on the $S^{d-1}$ by $r= L \sin \chi$ in the region $0 \leq \chi < \pi/2$. The range $\pi/2 < \chi \leq \pi$ covers the static patch of the antipodal observer. Note that, since $\tilde{r} = \cos \chi$, the relation (\ref{Rindlerobservers}) between $\partial_{X^0}$ and $\partial_t$ is reversed for this observer.

\begin{figure}[hbtp]
 \centering
  \epsfysize=6 cm
  \epsfbox{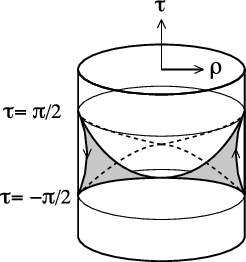}
 \centering
 \caption{Geometry of Rindler-AdS$_{d+1}$ space. The shaded region is a surface of constant $R$, which covers the static patches of a pair of antipodal de Sitter observers. $\tau$ and $\rho$ are the time and radius in global coordinates. The Rindler-AdS region extends only up to $\tau = \pm \pi/2$ at the boundary of AdS. The arrow in the right shaded region points in the direction of $\partial_t$, whose orbits are a Rindler/de Sitter observer's worldline; the arrow is reversed for the antipodal observer. Except at $\rho = 0$ each point in the interior corresponds to a $S^{d-2}$.}
 \label{fig:dS}
 \end{figure}

Incidentally, the spatial geometry at constant $t$ is given by
\be
ds^2 = {dR^2 \over 1 + (R/L)^2} + R^2 \lf d \chi^2 + \sin^2 \chi  d \Omega_{d-2}^2 \rt \; ,
\ee
which is locally Euclidean $AdS_d$ i.e. the hyperbolic space $H_d$. For the region $0 \leq \chi < \pi/2$ (corresponding to $0 \leq r < L$), the spatial part of AdS that corresponds to a Rindler observer is really $H_d /Z_2$ whose topology is $B^d /Z_2$. The geometry of Rindler-AdS space is depicted in Figure \ref{fig:dS}.

To compute the temperature of the Rindler horizon, consider a Rindler observer at constant $R$ and constant $r$. The proper acceleration of such an observer is
\be
a = {1 \over R} \sqrt{(R/L)^2 + {1 \over 1 - (r/ L)^2}} \; . \label{accel2}
\ee
Inserting (\ref{accel2}) into (\ref{tempgeneral}) we get
\be
T_{\rm local} = {1 \over 2 \pi R} \sqrt{1 \over 1 - (r/ L)^2} \; , \label{TphysDS}
\ee
and the horizon temperature is
\be
T_H = \sqrt{-g_{tt}}T_{\rm local} \; . \label{T-horizon}
\ee

From the boundary point of view the Rindler observer is an accelerating observer at fixed $r$ in static de Sitter space. To obtain the de Sitter temperature, we define $t = {\hat t}/(R/ L)^2$ which puts the constant $R$ part of the metric in the form
\be
ds^2 = - f(r) d{\hat t}^2 + {dr^2 \over f(r)} + \lf R/ L \rt ^2 r^2  d \Omega_{d-2}^2 \; .
\ee
Then the de Sitter temperature is
\be
T =  {f'(r_H) \over 4 \pi} = {l \over 2 \pi R^2} \; ,
\ee
and the local temperature at constant $r$ is
\be
T_{\rm boundary}= {1 \over 2 \pi R} {1 \over \sqrt{1-(r/ L)^2}} \; ,
\ee
which is again the physically-measured Rindler temperature.

The entropy of the Rindler horizon is calculated using the standard area formula. The horizon is at $r = \lm$. Specializing to $AdS_5$, the Rindler horizon has entropy
\be
S_{\rm Rindler} = {\pi \over G_5} \int_0^{R_0} \! \! {R^2 dR \over \sqrt {1 + (R/L)^2}} = {\pi L^2 \over 2 G_5} \lf R_0 \sqrt{1 + (R_0/L)^2} - L \sinh^{-1} (R_0/L) \rt \; ,
\ee
where $R_0$ is a cut-off radius which acts in the bulk as an infrared regulator. We see that for large $R_0$ the entropy scales like $R_0^2$:
\be
S_{\rm Rindler} \approx {\pi L R_0^2 \over 2 G_5} \; . \label{entropyscalingDeSitter}
\ee
The coordinate $R$ scales the boundary theory in this parameterization. At fixed $R = R_0$, therefore, the theory is a UV cut-off CFT in static de Sitter space. The $R_0^2$ scaling of the entropy, (\ref{entropyscalingDeSitter}), seems to indicate, perhaps surprisingly, that a free field computation for a thermal CFT will not give the right result either. A free field calculation, quite apart from being off by numerical factors, would be expected to yield an extensive entropy that scales like $R_0^3$ though oddly the entropy in this case is precisely $(R_0/L)^2 N^2$ using (\ref{gaugegroup}). The actual $R_0^2$ scaling strongly suggests that the correct boundary interpretation of Rindler entropy could be as entanglement entropy \cite{minimalsurfaces,bms,lohiya,Marolf}; the de Sitter horizon acts as a surface across which the conformal fields are entangled with the fields in the static de Sitter patch of the antipodal observer.  

To calculate the two-point correlator consider a massive scalar field in Rindler-AdS$_{d+1}$. The easiest way to calculate the boundary correlation functions is to Wick-rotate the time coordinate as $t \rightarrow i L \psi$;
the CFT then lives on an $S^d$. The two-point function of the dual operator can now be easily calculated as
\be 
 \langle  {\cal O} (\theta_1,\psi_1)  {\cal O} (\theta_2,\psi_2) \rangle = {1\over {\left (1 - \cos D \right)^\Delta}} \; , \label{2pointFndeSitter}
\ee
where $\Delta = 1 + \sqrt{1 + m^2}$, is the conformal dimension of the dual operator, and $D$ is the de Sitter invariant distance in $d$ dimensions, which in two dimensions becomes $\cos D = \left (\sin \theta_1 \sin \theta_2 \cos \left(\psi_1 - \psi_2\right) + \cos \theta_1 \cos \theta_2 \right)$. We observe that (\ref{2pointFndeSitter}) has the required periodicity in the imaginary time coordinate, $\psi$, and yields the correct Rindler temperature (\ref{T-horizon}). 

That a certain foliation of AdS has de Sitter space as its boundary is intriguing. It would be interesting to try to understand the vacuum states in de Sitter space \cite{Allen} using this setup. It may allow us to use the AdS/CFT correspondence in the reverse way: by using gravity in Rindler-AdS space to learn about strongly-coupled field theories in de Sitter space \cite{Marolf}. There are some subtleties, however. Unlike in our previous foliation, the boundary itself now contains a horizon, corresponding to the horizons of the static diamond of de Sitter space. The boundary horizon does not have finite entropy, however, since there is no gravity in the boundary theory.

\section{Summary and Discussion}
In this paper, we have presented a holographic duality for acceleration horizons. The key idea was to consider acceleration horizons in AdS, rather than in flat space, so as to be able to exploit the AdS/CFT correspondence. We then used the dual picture to holographically probe properties of the Rindler horizon. We recovered the horizon thermodynamics including the precise entropy density for the case of Rindler-AdS$_3$. We also showed that physics beyond the horizon can be probed from the perspective of the boundary theory by calculating the response of the boundary theory to an infalling horizon-crossing source. Evidently, Rindler-AdS/CFT holds much promise for studying the quantum gravity of horizons and, moreover, it is considerably more tractable than the  holography of AdS-Schwarzschild black holes; we have surely only scratched the surface of this rich subject.

Among the obvious directions for future study are to work out two-point and higher correlation functions for infalling sources and to look at other more realistic scenarios that might probe the horizon. It would be particularly interesting to set up a problem in which information fell into the Rindler horizon, to see whether our intuition about information return is borne out. Another obvious direction is to perform calculations using Rindler-AdS/CFT and then finally make a global identification in the $\chi$ direction to learn about the holography of BTZ black holes.

Also, as mentioned earlier, there are subtleties in the Lorentzian version of AdS/CFT because of the presence of normalizable modes. We ignored in this work but it would be interesting to work out mode solutions for (\ref{RindlerAdSmetric}) and map them to the boundary theory. One can also determine the spectrum of normalizable modes and study the quantization conditions. This will throw more light on the dictionary between the bulk and the boundary descriptions in Rindler-AdS/CFT.

It should be noted that what we have done was, in some sense, still quantum field theory in curved spacetime. The boundary theory learned about the bulk from the boundary value of the bulk field which in turn was determined using a propagator over a fixed background geometry. By considering graviton fluctuations, we might be able to take a step beyond QFT in curved spacetime.

More speculatively, we could try to implement some kind of observer complementarity \cite{complementarity,elliptic}. For example, in our scenario we know that complete information about the switching off event in the upper Rindler wedge was provided by the pole structure in both CFTs. In order for all this information to be available to one observer, it might be necessary to perform some kind of antipodal identification \cite{elliptic} or to map the antipodal CFT to some other surface in the original wedge, such as at the stretched horizon \cite{complementarity,mempar}. It might also be, however, that complete information is not provided even by both CFTs. In particular, the points where the two antipodal Rindler wedges intersect cannot be attributed unambiguously to either Rindler wedge. Correspondingly, operator insertions on the boundary of global AdS at precisely the points where it touches that intersection surface cannot obviously be thought of as insertions in either of the two CFTs.

Still more speculatively, there might be connections to the Hagedorn transition. In quantum field theory, acceleration and temperature are linearly related, but in string theory it is possible that something nontrivial happens when the temperature reaches the Hagedorn temperature. Perhaps the existence of a Rindler-AdS/CFT correspondence might provide a new angle from which to examine this old issue.

\bigskip
\noindent
{\bf Acknowledgments}

\noindent
We thank Erik Verlinde for collaborating on this project during its early stages, as well as for subsequent discussions. We would also like to thank Jan de Boer, Paul Davies, and David Lowe for helpful discussions. P. S. would like to thank Yao Ji and Thomas Jacques for their generous help in making figures. M. P. is supported in part by John Templeton Foundation grant 60253.

\section*{Appendix}
\subsection*{Rindler coordinates for AdS$_{d+1}$}
To view Rindler observers as part of AdS, define
\begin{eqnarray}
X^0 &=& {\xi} \sinh (t/ L) \nonumber \\
X^1 &=& {\xi} \cosh (t/ L) \nonumber \\
X^2 & = & \sqrt{L^2 + \xi^2} \sinh (\chi / L) \cos \theta_1  \nonumber \\
& ... &  \nonumber \\
X^{d-2} & = &\sqrt{L^2 + \xi^2} \sinh (\chi / L) \sin \theta_1 ... \sin \theta_{d-3} \cos \theta_{d-2} \nonumber \\
X^{d-1} & = &\sqrt{L^2 + \xi^2} \sinh (\chi / L) \sin \theta_1 ... \sin \theta_{d-2} \cos \phi \nonumber \\
X^d & = &\sqrt{L^2 + \xi^2} \sinh (\chi / L) \sin \theta_1 ... \sin \theta_{d-2} \sin \phi \nonumber \\
X^{d+1} & = & \sqrt{L^2 + \xi^2} \cosh (\chi / L) \; . \label{Rindlercoords}
\end{eqnarray}
This satisfies the AdS embedding equation (\ref{embed}). 
The ranges of the coordinates are
\be
0 < \xi  \qquad -\infty < t < \infty \qquad -\infty < \chi <
\infty\qquad 0 \leq \theta_i \leq \pi \qquad 0 \leq \phi < 2 \pi \; .
\ee

\subsection*{Global coordinates for AdS$_{d+1}$}
Global coordinates are related to embedding coordinates via
\begin{eqnarray}
X^0 & = & \sqrt{L^2 + \rho^2} \sin (\tau/L) \nonumber \\
X^1 & = & \rho \cos \psi  \nonumber \\
X^2 & = & \rho \sin \psi \cos \theta_1  \nonumber \\
& ... &  \nonumber \\
X^{d-2} & = & \rho \sin \psi \sin \theta_1 ... \sin \theta_{d-3} \cos \theta_{d-2} \nonumber \\
X^{d-1} & = & \rho \sin \psi \sin \theta_1 ... \sin \theta_{d-2} \cos \phi \nonumber \\
X^d & = & \rho \sin \psi \sin \theta_1 ... \sin \theta_{d-2} \sin \phi \nonumber \\
X^{d+1} & = & \sqrt{L^2 + R^2} \cos (\tau/L) \; .
\end{eqnarray}

\subsection*{Eddington-Finkelstein coordinates for Rindler-AdS$_3$}
Ingoing Eddington-Finkelstein coordinates are related to AdS embedding coordinates through
\begin{eqnarray}
X^0 &=& \sqrt{2rL} \sinh \left ( \frac{1}{L}(v - f( r)) \right ) \nonumber \\
&=& \f{1}{2}\left[e^{v/L} \sqrt{\f{2 r L(\sqrt{1 + 2r/L} + 1)}{\sqrt{1 + 2r/L} - 1}} - e^{-v/L} \sqrt{\f{2 r L(\sqrt{1 + 2r/L} - 1)}{\sqrt{1 + 2r/L} + 1}}\right] \nonumber \\
\\
X^1 &=& \sqrt{2rL} \cosh \left ( \frac{1}{L}(v - f( r)) \right ) \nonumber \\
&=& \f{1}{2}\left[e^{v/L} \sqrt{\f{2 r L(\sqrt{1 + 2r/L} + 1)}{\sqrt{1 + 2r/L} - 1}} + e^{-v/L} \sqrt{\f{2 r L(\sqrt{1 + 2r/L} - 1)}{\sqrt{1 + 2r/L} + 1}}\right]  \nonumber \\
\\
X^2 &=& \sqrt{L^2 + 2rL} \sinh \left (\frac{\chi}{L} \right) \nonumber \\
X^3 &=& \sqrt{L^2 + 2rL} \cosh \left (\frac{\chi}{L}\right ) \; ,
\end{eqnarray}
where $f(r) = \f{L}{2}\ln \left[\f{\sqrt{1 + \f{2r}{L}} - 1}{\sqrt{1 + \f{2r}{L}} + 1} \right]$ as given by (\ref{Eddingtontime}). These coordinates are nonsingular at the Rindler horizon $r = 0$.

\end{document}